\documentclass[twocolumn,showpacs,preprintnumbers,amsmath,amssymb]{revtex4}

\usepackage{graphicx}
\usepackage{dcolumn}
\usepackage{bm}
\usepackage{epsfig}

\def\be{\begin{eqnarray}}
\def\en{\end{eqnarray}}
\def\non{\nonumber}
\def\la{\langle}
\def\ra{\rangle}

\def\vp{\varepsilon}

\def\B{{\cal B}}

\def\ov{\overline}
\def\pr{{Phys. Rev.}~}
\def\prl{{ Phys. Rev. Lett.}~}
\def\pl{{ Phys. Lett.}~}
\def\np{{ Nucl. Phys.}~}
\def\zp{{ Z. Phys.}~}
\def\lsim{ {\ \lower-1.2pt\vbox{\hbox{\rlap{$<$}\lower5pt\vbox{\hbox{$\sim$}
}}}\ } }
\def\gsim{ {\ \lower-1.2pt\vbox{\hbox{\rlap{$>$}\lower5pt\vbox{\hbox{$\sim$}
}}}\ } }

\begin{document}


\title{Status of Exclusive Baryonic $B$ Decays}

\author{Hai-Yang Cheng}

\affiliation{%
Institute of Physics, Academia Sinica, Taipei, Taiwan 115,
Republic of China
}%

\begin{abstract}
The status of exclusive two-body and three-body baryonic $B$
decays is reviewed. The threshold peaking effect in baryon pair
invariant mass is stressed and explained. Weak radiative baryonic
$B$ decays mediated by the electromagnetic penguin process are
discussed.

\end{abstract}

\maketitle

\section{Introduction}

A unique feature of hadronic $B$ decays is that the $B$ meson is
heavy enough to allow a baryon-antibaryon pair production in the
final state. During the Lepton-Photon Conference in 1987, ARGUS
announced the first measurement of the decay modes $p\bar
p\pi^\pm$ and $p\bar p\pi^+\pi^-$ in $B$ decays at the level of
$10^{-4}$ \cite{ARGUS}. Although this observation of charmless
baryonic $B$ decays was immediately ruled out by CLEO
\cite{CLEO89}, it nevertheless has stimulated extensive
theoretical studies during the period of 1988-1992. Several
different model frameworks have been proposed: the constituent
quark model \cite{Korner}, the pole model \cite{DTS,Jarfi}, the
QCD sum rule \cite{Chernyak}, the diquark model \cite{Ball} and
flavor symmetry considerations \cite{Gronau}.

However, experimental and theoretical activities in baryonic $B$
decays suddenly faded away after 1992. This situation was
dramatically changed in the past three years. Interest in this
area was revived by many new measurements at CLEO and Belle
followed by several theoretical studies.

\subsection{Experimental status}
\noindent{\it two-body decays}:
Except for the recently measured
$\ov B^0\to \Lambda_c^+\bar p$ by Belle \cite{Lamcp}
  \be
 \B(\ov B^0\to\Lambda_c^+\bar p) &=& (2.19^{+0.56}_{-0.49}\pm
 0.32\pm0.57)\times 10^{-5},
 \en
none of the two-body baryonic $B$ decays has been observed. The
experimental upper limits are summarized in Tables \ref{tab:2body}
and \ref{tab:3charm} for charmless and charmful decays,
respectively. We see that the present limit on charmless ones is
generally of order $10^{-6}$ except for the $p\bar p$ mode which
was recently pushed down to the level of $2.7\times 10^{-7}$ by
BaBar \cite{BaBar:pp}.

\begin{table}[h]
\caption{Experimental upper limits on the branching ratios of
charmless two-body baryonic $B$ decays. } \label{tab:2body}
\begin{ruledtabular}
\begin{tabular}{l l l l  }
 Decay & CLEO \cite{CLEO:2body} & Belle \cite{Belle:2body} & BaBar \cite{BaBar:pp} \\ \hline
 $\ov B^0\to p\bar p$ & $1.4\times 10^{-6}$ & $1.2\times
 10^{-6}$ & $2.7\times 10^{-7}$ \\
 $\ov B^0\to\Lambda\bar\Lambda$ & $1.2\times 10^{-6}$ & $1.0\times 10^{-6}$ \\
 $B^-\to\Lambda\bar p$ & $1.5\times 10^{-6}$ & $2.2\times
 10^{-6}$ \\
 $B^-\to p\bar\Delta^{--}$ & $1.5\times 10^{-4}$ & \\
 $B^-\to\Delta^0\bar p$ & $3.8\times 10^{-4}$ & \\
 $\ov B^0\to\Delta^{++}\Delta^{--}$ & $1.1\times 10^{-4}$ & \\
 $\ov B^0\to \Delta^0\bar\Delta^0$ & $1.5\times 10^{-3}$ & \\
\end{tabular}
\end{ruledtabular}
\end{table}

\vspace{0.2cm} \noindent{\it three-body decays}: Unlike the
two-body case, the measurements of three-body or four-body
baryonic $B$ decays are quite fruitful and many new results have
been emerged in recent years. For the charmless case, Belle
\cite{Belle:3charmless} has observed 5 modes, see Table
\ref{tab:3charmless}. The channel $B^-\to p\bar p K^-$ announced
by Belle nearly two years ago \cite{Belle:ppK} is the first
observation of charmless baryonic $B$ decay.

\begin{table}[h]
\caption{Branching ratios of charmless three-body baryonic $B$
decays measured by Belle \cite{Belle:3charmless}. }
\label{tab:3charmless}
\begin{ruledtabular}
\begin{tabular}{l l l   }
 Decay & Br($10^{-6}$)  \\ \hline
 $\ov B^0\to\Lambda\bar p\pi^+$ &
 $3.97^{+1.00}_{-0.80}\pm0.56$ \\
 $B^-\to p\bar p K^-$ & $5.66^{+0.67}_{-0.57}\pm0.62$ \\
 $B^-\to p\bar p K^{*-}$ & $10.3^{+3.6+1.3}_{-2.8-1.7}$ \\
 $\ov B^0\to p\bar p K_S$  & $1.88^{+0.77}_{-0.60}\pm0.23$ \\
 $B^-\to p\bar p\pi^-$ & $3.06^{+0.73}_{-0.62}\pm0.37$ \\
 $\ov B^0\to\Lambda \bar pK^+$ & $<0.82$ \\
 $\ov B^0\to\Sigma^0\bar p\pi^+$ & $<3.8$ \\
\end{tabular}
\end{ruledtabular}
\end{table}

Table \ref{tab:3charm} summarizes the measured branching ratios of
charmful baryonic decays with a charmed meson or a charmed baryon
in the final state. In general, Belle \cite{Belle:3charm} and CLEO
\cite{CLEO:3charm} results are consistent with each other except
for the ratio of $\Sigma_c^{++}\bar p\pi^-$ to $\Sigma_c^0\bar
p\pi^+$. The $\Sigma_c^{++}$ decay proceeds via both external and
internal $W$-emission diagrams, whereas the $\Sigma_c^0$ decay can
only proceed via an internal $W$ emission. While Belle
measurements imply a sizable suppression for the $\Sigma_c^0$
decay (and likewise for the $\Sigma_{c1}$ decay), it is found by
CLEO that $\Sigma_c^{++}\bar p\pi^-$, $\Sigma_c^0\bar p\pi^+$ and
$\Sigma_c^0\bar p\pi^0$ are of the same order of magnitude.
Therefore, it is concluded by CLEO that the external $W$ decay
diagram does not dominate over the internal $W$-emission diagram
in Cabibbo-allowed baryonic $B$ decays. This needs to be clarified
by the forthcoming improved measurements. The decay $B^-\to
J/\psi\Lambda\bar p$ was recently measured by BaBar with the
branching ratio $(12^{+9}_{-6})\times 10^{-6}$ \cite{BaBar:JLamp}
and an upper limit $4.1\times 10^{-5}$ was set by Belle
\cite{Belle:JLamp}.

\begin{table*}
\caption{Experimental measurements of the branching ratios (in
units of $10^{-4}$) for the $B$ decay modes with a charmed baryon
$\Lambda_c$ or $\Lambda_{c1}=\Lambda_c(2593),\Lambda_c(2625)$ or
$\Sigma_c(2455)$ or $\Sigma_{c1}=\Sigma_c(2520)$ or a charmed
meson in the final state.
 } \label{tab:3charm}
\begin{ruledtabular}
\begin{tabular}{l l l   }
 Mode~~ & Belle \cite{Belle:3charm,Lamcp} & CLEO  \cite{CLEO:3charm} \\ \hline
  $B^-\to\Lambda_c^+\bar p\pi^-\pi^0$ & &
 $18.1\pm2.9^{+2.2}_{-1.6}\pm4.7$ \\
 $\ov B^0\to\Lambda_c^+\bar p\pi^+\pi^-$ &
 $11.0\pm1.2\pm1.9\pm 2.9$ &
 $16.7\pm1.9^{+1.9}_{-1.6}\pm4.3$  \\
 $B^-\to \Lambda_c^+\bar p\pi^-$ & $1.87^{+0.43}_{-0.40}\pm0.28\pm
 0.49$ & $2.4\pm0.6^{+0.19}_{-0.17}\pm0.6$  \\
 $\ov B^0\to\Lambda_c^+\bar p$ &
 $0.22^{+0.06}_{-0.05}\pm0.03\pm0.06$ & $<0.9$  \\
 $B^-\to\Lambda_{c1}^+\bar p\pi^-$ & & $<1.9$ \\
 $\ov B^0\to\Lambda_{c1}^+\bar p$ & & $<1.1$ \\ \hline
 $B^-\to \Sigma_c^{++}\bar p\pi^-\pi^-$ & & $2.8\pm 0.9\pm 0.5\pm
 0.7$  \\
 $B^-\to\Sigma_c^0\bar p\pi^+\pi^-$ & & $4.4\pm1.2\pm0.5\pm1.1$ \\
 $\ov B^0\to \Sigma_c^{++}\bar p\pi^-$ &
 $2.38^{+0.63}_{-0.55}\pm0.41\pm0.62$ & $3.7\pm0.8\pm0.7\pm0.8$
 \\
 $\ov B^0\to\Sigma_c^0\bar p\pi^+$ &
 $0.84^{+0.42}_{-0.35}\pm0.14\pm0.22<1.59$ & $ 2.2\pm0.6\pm0.4\pm0.5$
 \\
 $B^-\to\Sigma_c^0\bar p\pi^0$ & & $4.2\pm 1.3\pm0.4\pm1.0$ \\
 $B^-\to\Sigma_c^0\bar p$ & $0.45^{+0.26}_{-0.19}\pm0.07\pm0.12<0.93$ &
 $<0.8$  \\
 $\ov B^0\to \Sigma_{c1}^{++}\bar p\pi^-$ &
 $1.63^{+0.57}_{-0.51}\pm0.28\pm0.42$ & \\
 $\ov B^0\to\Sigma_{c1}^0\bar p\pi^+$ &
 $0.48^{+0.45}_{-0.40}\pm0.08\pm0.12<1.21$ &  \\
 $B^-\to\Sigma_{c1}^0\bar p$ & $0.14^{+0.15}_{-0.09}\pm0.02\pm0.04<0.46$ &  \\
 \hline
 $\ov B^0\to D^{*+}n\bar p$ & & $14.5^{+3.4}_{-3.0}\pm2.7$ \\
 $\ov B^0\to D^0 p\bar p$ & $1.18\pm0.15\pm0.16$ & \\
 $\ov B^0\to D^{*0} p\bar p$ & $1.20^{+0.33}_{-0.29}\pm0.21$ & \\
\end{tabular}
\end{ruledtabular}
\end{table*}

\begin{figure}[h]
\includegraphics[width=5.0cm]{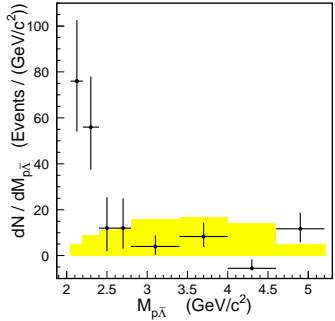}
\caption{The $p\bar\Lambda$ invariant mass distribution of $B^0\to
\bar \Lambda p\pi^-$ \cite{Belle:3charmless}.}
\label{fig:spectrum}
\end{figure}

A common and unique feature of the spectrum for $B\to \B_1\ov \B_2
M$ (e.g. $\ov B^0\to\Lambda\bar p\pi^+$) is the threshold
enhancement behavior of the baryon-pair invariant mass (see Fig.
\ref{fig:spectrum}): It sharply peaks at very low values. That is,
the $B$ meson is preferred to decay into a baryon-antibaryon pair
with low invariant mass accompanied by a fast recoil meson.

Now it is well established experimentally that
 \be
 \B(B^-\to\Lambda_c^+\bar p\pi^-) &\gg& \B(\ov B^0\to\Lambda_c^+\bar
 p), \non \\
 \B(B^-\to p\bar p K^-) &\gg&  \B(\ov B^0\to p\bar p), \non\\
 \B(B^-\to\Sigma_c^0\bar p\pi^0) &\gg& \B(B^-\to\Sigma_c^0\bar p).
 \en
Therefore, some three-body final states have rates larger than
their two-body counterparts. This phenomenon can be understood in
terms of the threshold effect, namely, the invariant mass of the
baryon pair is preferred to be close to the threshold. The
configuration of the  two-body decay $B\to\B_1\ov \B_2$ is not
favorable since its invariant mass is $m_B$. In $B\to \B_1\ov\B_2
M$ decays, the effective mass of the baryon pair is reduced as the
emitted meson can carry away much energies.

\subsection{Theoretical progress}
Since baryonic $B$ decays involve two baryons, it is extremely
complicated and much involved. Nevertheless, there are some
theoretical progresses in the past three years.

It is known that two-body baryonic $B$ decays are dominated by
nonfactorizable contributions that are difficult to evaluate. This
nonfactorizable effect can be evaluated in the pole model. Using
the MIT bag model to evaluate the weak matrix elements and the
$^3P_1$ model to estimate the strong coupling constants, it is
found in \cite{CKcharmless} and \cite{CKcharm} that the charmless
and charmful two-body decays can be well described. Chang and Hou
\cite{Chang} have generalized the original version of the diquark
model in \cite{Ball} to include penguin effects, but no
quantitative predictions have been made. In the past year, a
diagrammatic approach has been developed for both charmful
\cite{Luo} and charmless \cite{Chua2body} decays.

\begin{figure*}
\centerline{
            {\epsfxsize2.3 in \epsffile{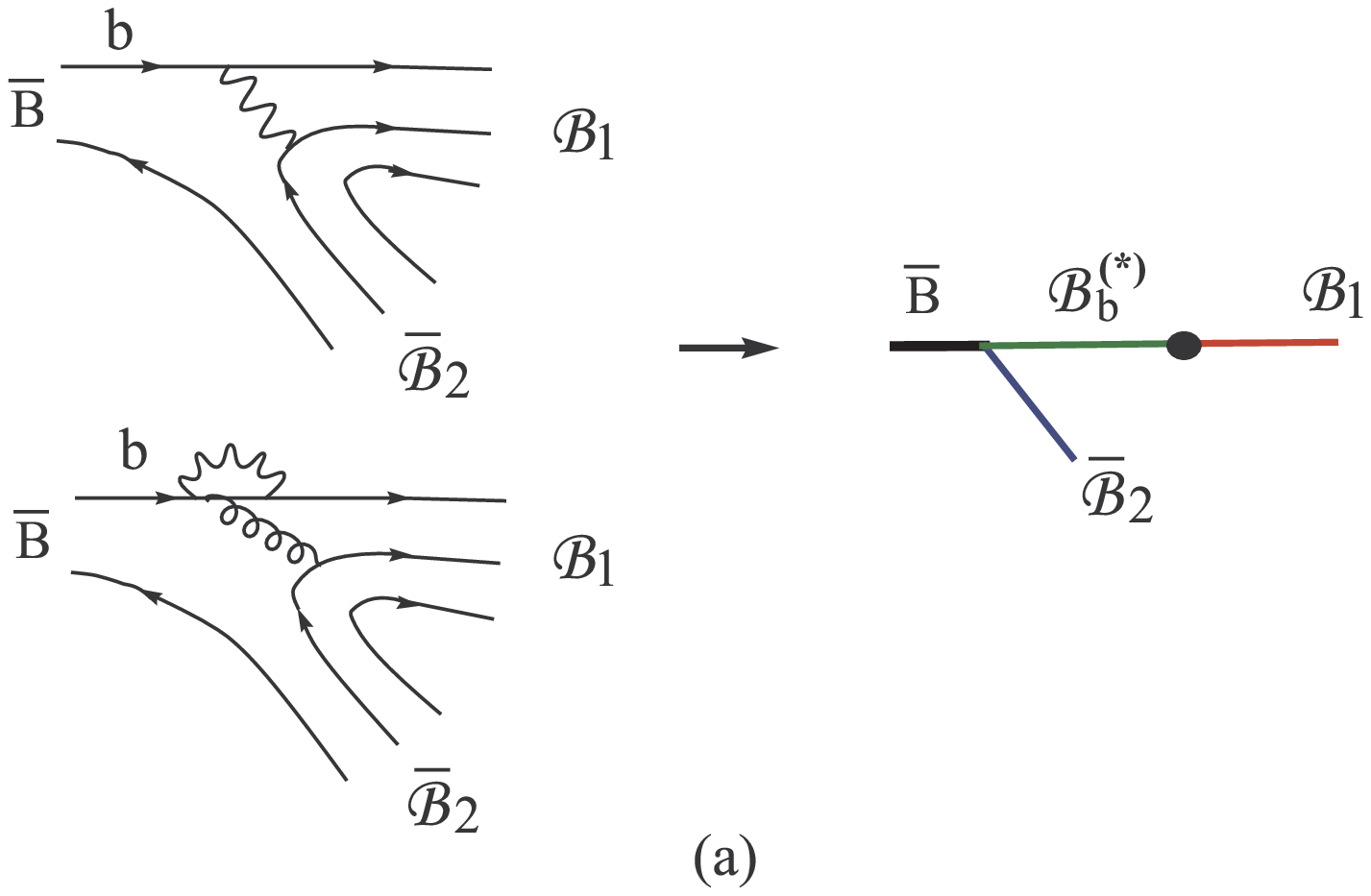}}
            {\epsfxsize2.3 in \epsffile{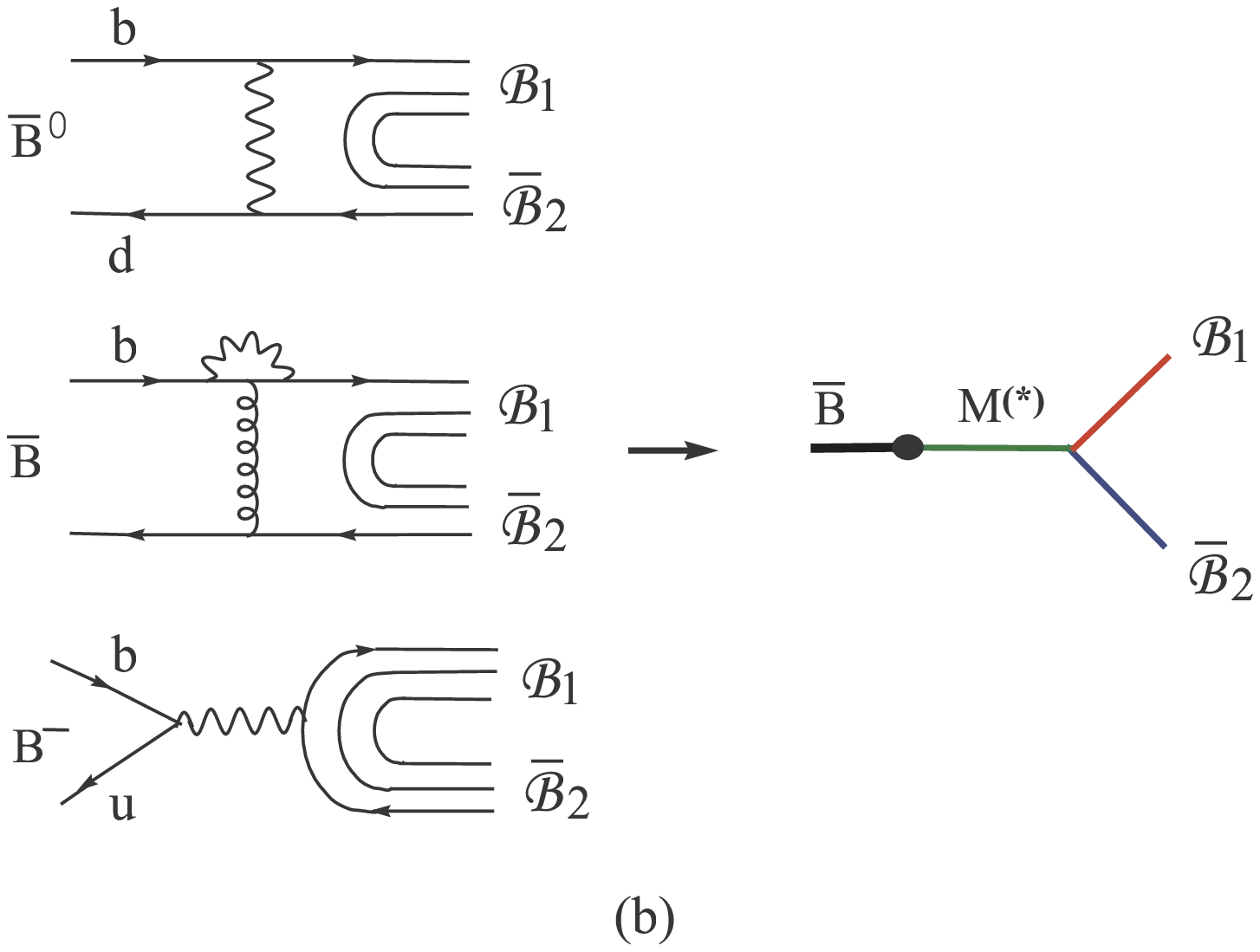}}}
\caption{Quark and pole diagrams for two-body baryonic $B$ decay
    $\ov B\to \B_1\ov \B_2$, where the symbol $\bullet$ denotes the weak
    vertex.} \label{fig:2body}
\end{figure*}

As pointed out by Dunietz \cite{Dunietz} and by Hou and Soni
\cite{HS}, the smallness of the two-body baryonic decay
$B\to\B_1\ov\B_2$ has to do with a straightforward Dalitz plot
analysis  or with the large energy release. Hou and Soni
conjectured that in order to have larger baryonic $B$ decays, one
has to reduce the energy release and  at the same time allow for
baryonic ingredients to be present in the final state.  This is
indeed the near threshold effect mentioned before. Of course, one
has to understand the underlying origin of the threshold peaking
effect.

Contrary to the two-body baryonic $B$ decay, the three-body decays
do receive factorizable contributions that fall into two
categories: (i) the transition process with a meson emission, $\la
M|(\bar q_3 q_2)|0\ra\la \B_1\ov \B_2|(\bar q_1b)|B\ra$, and (ii)
the current-induced process governed by the factorizable amplitude
$\la \B_1\ov \B_2|(\bar q_1 q_2)|0\ra \la M|(\bar q_3 b)|B\ra$.
The two-body matrix element $\la \B_1\ov \B_2|(\bar q_1 q_2)|0\ra$
in the latter process can be either related to some measurable
quantities or calculated using the quark model. The
current-induced contribution to three-body baryonic $B$ decays has
been discussed in various publications \cite{CHT01,CHT02,CH02}. On
the contrary, it is difficult to evaluate the three-body matrix
element in the transition process and in this case one can appeal
to the pole model \cite{CKcharmless,CKcharm,CKDmeson}.

Weak radiative baryonic $B$ decays $B\to\B_1\ov \B_2\gamma$
mediated by the electromagnetic penguin process $b\to s\gamma$ may
have appreciable rates. Based on the pole model, it is found that
$B^-\to\Lambda\bar p\gamma$ and $B^-\to\Xi^0\bar\Sigma^-\gamma$
have sizable rates and are readily accessile \cite{CKrad}.

\section{2-Body baryonic $B$ decays}
As shown in Fig. 2, the quark diagrams for two-body baryonic $B$
decays consist of  internal $W$-emission diagram, $b\to d(s)$
penguin transition, $W$-exchange for the neutral $B$ meson and
$W$-annihilation for the charged $B$. Just as mesonic $B$ decays,
$W$-exchange and $W$-annihilation are expected to be helicity
suppressed and the former is furthermore subject to color
suppression. Therefore, the two-body baryonic $B$ decay
$B\to\B_1\ov \B_2$ receives the main contributions from the
internal $W$-emission diagram for tree-dominated modes and the
penguin diagram for penguin-dominated processes.

\begin{figure*}[t]
\centerline{
            {\epsfxsize3.0 in \epsffile{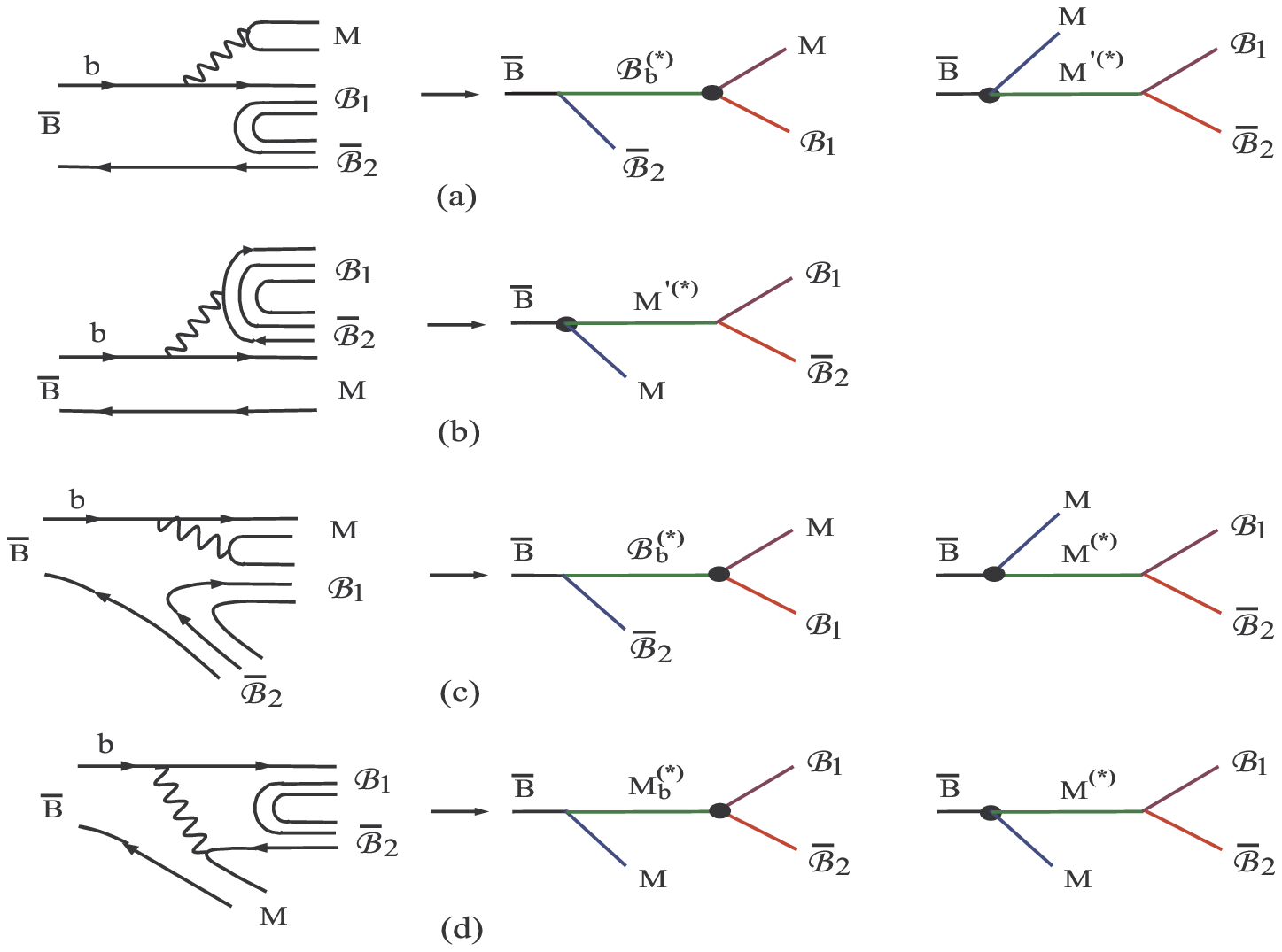}}
            {\epsfxsize2.3 in \epsffile{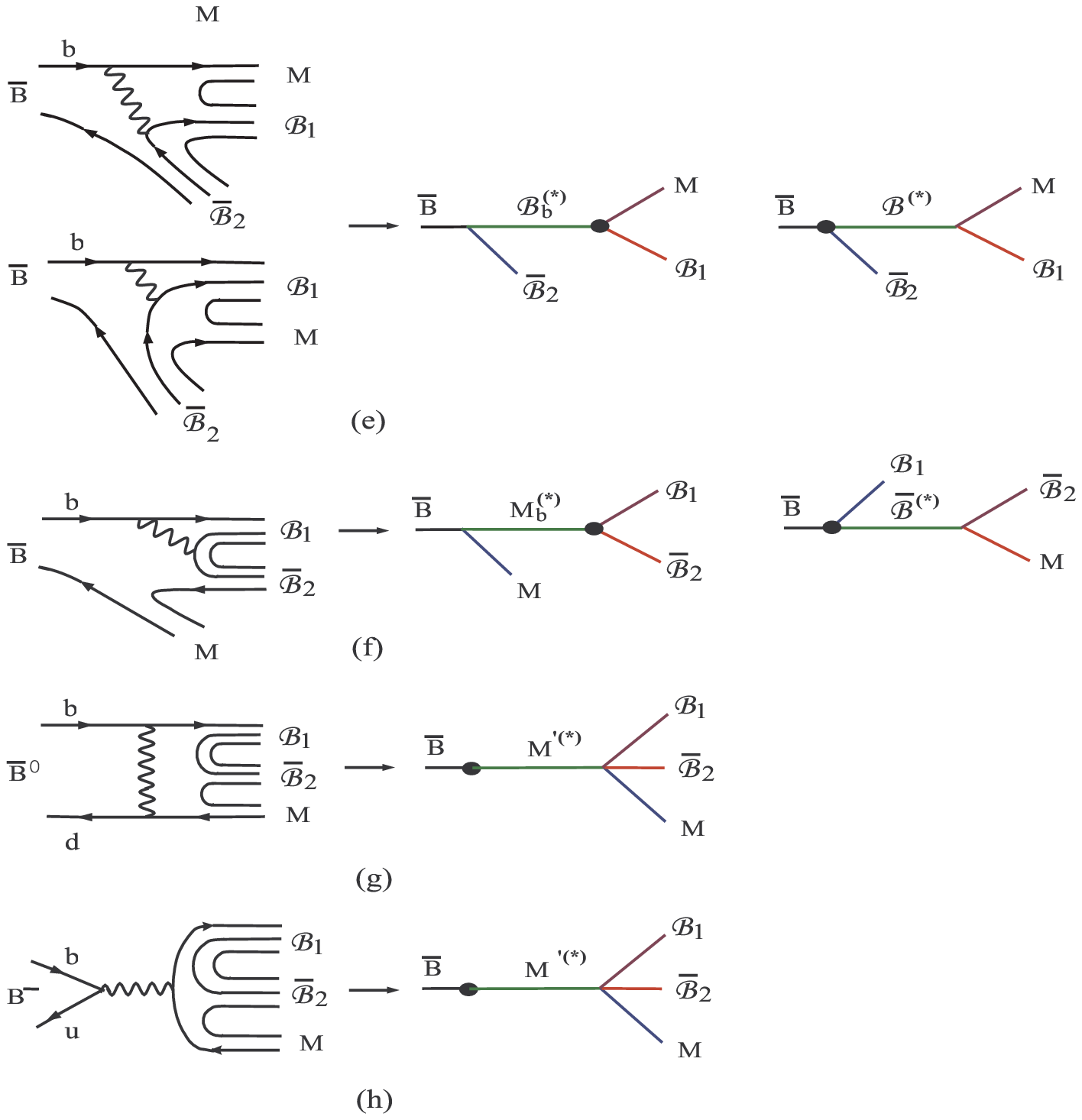}}}
    \caption{ Quark and pole diagrams for three-body baryonic $B$ decay
    $\ov B\to \B_1\ov \B_2M$, where the symbol $\bullet$ denotes the weak
    vertex.}
\end{figure*}

These amplitudes are nonfactorizable in nature and thus very
difficult to evaluate directly.  In order to circumvent this
difficulty, it is customary to assume that the decay amplitude at
the hadron level is dominated by the pole diagrams with low-lying
one-particle intermediate states. The general amplitude reads
  \be
 {\cal A}(B\to \B_1\ov \B_2)=\bar u_1(A+B\gamma_5)v_2,
 \en
where $A$ and $B$ correspond to $p$-wave parity-violating (PV) and
$s$-wave parity-conserving (PC) amplitudes, respectively. In the
pole model, PC and PV amplitudes are dominated by ${1\over 2}^+$
ground-state intermediate states and ${1\over 2}^-$ low-lying
baryon resonances, respectively.  This pole model has been applied
successfully to nonleptonic decays of hyperons and charmed baryons
\cite{CT92,CT93}. In general, the pole diagram leads to
 \be \label{eq:AandB}
 A=-\sum_{\B_b^*}{g_{\B_b^{*}\to B\B_2}\,b_{\B_b^*\B_1}\over
 m_{1}-m_{\B_b^*} }, ~~ B=\sum_{\B_b}{g_{\B_b\to
 B\B_2}\,
 a_{\B_b\B_1}\over m_{1}-m_{\B_b}}.
 \en

There are two unknown quantities in the above equation: weak
matrix elements and strong couplings. For the former we employ the
MIT bag model to evaluate the baryon-to-baryon transitions
\cite{CKcharmless}. For the latter, there are two distinct models
for quark pair creation: (i) the $^3P_0$ model in which the $q\bar
q$ pair is created from the vacuum with vacuum quantum numbers.
Presumably it works in the nonperturbative low energy regime, and
(ii) the $^3S_1$ model in which the quark pair is created
perturbatively via one gluon exchange with one-gluon quantum
numbers $^3S_1$. Since the light baryons produced in two-body
baryonic $B$ decays are very energetic, it appears that the
$^3S_1$ model may be more relevant.

\begin{table}[h]
\caption{Predictions of charmful two-body baryonic $B$ decays.
Experimental results are taken from Table \ref{tab:3charm}.}
\label{tab:charm2body}
\begin{ruledtabular}
\begin{tabular}{l c c c  } & \cite{Jarfi} & \cite{CKcharm} & Expt.
\\ \hline
 $\ov B^0\to\Lambda_c^+\bar p$ & $1.1\times 10^{-3}$
 & $1.1\times 10^{-5}$ & $(2.19\pm0.84)\times 10^{-5}$ \\
 $B^-\to\Sigma_c^0\bar p$ & $1.5\times 10^{-2}$ & $6.0\times 10^{-5}$ & $< 8.0\times 10^{-5}$ \\
 $\ov B^0\to\Sigma_c^0\bar n$ & $5.8\times 10^{-3}$ & $6.0\times 10^{-7}$ & \\
 $B^-\to\Lambda_c^+\bar\Delta^{--}$ & $3.6\times 10^{-2}$ & $1.9\times 10^{-5}$ & \\
\end{tabular}
\end{ruledtabular}
\end{table}

The predictions for charmful $B\to \B_1\ov\B_2$ decays are
summarized in Table \ref{tab:charm2body}. All earlier predictions
based on the sum-rule analysis, the pole model and the diquark
model are too large compared to experiment. Note that we predict
that $B^-\to \Sigma_c^0\bar p$ has a larger rate than $\ov B^0\to
\Lambda_c\bar p$ since the former proceeds via the $\Lambda_b$
pole while the latter via $\Sigma_b$ pole and the $\Lambda_b N\bar
B$ coupling is larger than $\Sigma_bN\bar B$ \cite{CKcharm}.
Therefore, it is important to measure the $\Sigma_c^0\bar p$
production to test the pole model.

From Table \ref{tab:charmless2body} we see that the charmless
two-body baryonic decays are predicted at the level of $10^{-7}$.
This is consistent with the observation of $\ov
B^0\to\Lambda_c^+\bar p$ after a simple scaling of
$|V_{ub}/V_{cb}|^2$.

\begin{table}[h]
\caption{Predictions of the branching ratios for some charmless
two-body baryonic $B$ decays classified into two categories:
tree-dominated and penguin-dominated. Branching ratios denoted by
``$\dagger$" are calculated only for the parity-conserving part.
Experimental limits are taken from Table \ref{tab:2body}.}
\label{tab:charmless2body}
\begin{ruledtabular}
\begin{tabular}{l c c c c  } & \cite{Chernyak} & \cite{Jarfi} & \cite{CKcharmless} & Expt. \\
\hline
 $\ov B^0\to p\bar p$ & $1.2\times 10^{-6}$ & $7.0\times
 10^{-6}$  & $1.1\times 10^{-7\dagger}$
 & $<2.7\times  10^{-7}$ \\
 $\ov B^0\to n\bar n$ & $3.5\times 10^{-7}$ & $7.0\times
 10^{-6}$ &  $1.2\times
 10^{-7\dagger}$ & \\
 $B^-\to n\bar p$ &  $6.9\times 10^{-7}$ & $1.7\times
 10^{-5}$ & $5.0\times  10^{-7}$ & \\
 $\ov B^0\to\Lambda\bar\Lambda$ &  & $2\times 10^{-7}$ & $0^\dagger$ &
 $<1.0\times 10^{-6}$ \\
 $B^-\to p\bar \Delta^{--}$ &  $2.9\times 10^{-7}$
 &  $3.2\times 10^{-4}$
 &  $1.4\times 10^{-6}$ & $<1.5\times 10^{-4}$ \\
 $\ov B^0\to p\bar\Delta^-$ &  $7\times 10^{-8}$ & $1.0\times
 10^{-4}$ & $4.3\times 10^{-7}$ & \\
 $B^-\to n\bar\Delta^-$ &
 & $1\times 10^{-7}$ & $4.6\times  10^{-7}$ & \\
 $\ov B^0\to n\bar\Delta^0 $ &  & $1.0\times 10^{-4}$ & $4.3\times 10^{-7}$ & \\
 \hline
 $B^-\to\Lambda\bar p$ &  $\lsim 3\times 10^{-6}$ & & $2.2\times 10^{-7\dagger}$ &
 $<1.5\times 10^{-6}$ \\
 $\ov B^0\to \Lambda\bar n$ & & & $2.1\times 10^{-7\dagger}$ & \\
 $\ov B^0\to\Sigma^+\bar p$ &   $6\times 10^{-6}$ & &  $1.8\times
 10^{-8\dagger}$ & \\
 $B^-\to\Sigma^0\bar p$ &  $3\times 10^{-6}$ & &  $5.8\times 10^{-8\dagger}$ & \\
 $B^-\to\Sigma^+\bar\Delta^{--}$ &  $6\times 10^{-6}$ & &  $2.0\times 10^{-7}$ & \\
 $\ov B^0\to\Sigma^+\bar\Delta^-$ &  $6\times 10^{-6}$ & &  $6.3\times 10^{-8}$ & \\
 $B^-\to\Sigma^-\bar\Delta^0$ &  $2\times 10^{-6}$ & &  $8.7\times 10^{-8}$ & \\
\end{tabular}
\end{ruledtabular}
\end{table}

\section{3-Body baryonic $B$ decays}

In the three-body baryonic decay, the emission of the meson $M$
will carry away energies in such a way that the invariant mass of
$\B_1\ov\B_2$ becomes smaller and hence it is relatively easier to
fragment into the baryon-antibaryon pair. One can also understand
this feature more concretely by studying the Dalitz plot. Due to
the $V-A$ nature of the $b\to ud\bar u$ process, the invariant
mass of the diquark $ud$ peaks at the highest possible values in a
Dalitz plot for $b\to ud\bar d$ transition \cite{Buchalla}. If the
$ud$ forms a nucleon, then the very massive $udq$ objects will
intend to form a highly excited baryon state such as $\Delta$ and
$N^*$ and will be seen as $N n\pi(n\geq 1)$ \cite{Dunietz}. This
explains the non-observation of the $N\ov N$ final states and why
the three-body mode $N\ov N \pi(\rho)$ is favored. Of course, this
argument is applicable only to the tree-dominated processes.

The quark diagrams and the corresponding pole diagrams for decays
of $B$ mesons to the baryonic final state $\B_1\ov\B_2 M$ are more
complicated. In general there are two external $W$-diagrams Figs.
3(a)-3(b), four internal $W$-emissions Figs. 3(c)-3(f), and one
$W$-exchange Fig. 3(g) for the neutral $B$ meson and one
$W$-annihilation Fig. 3(h) for the charged $B$. Because of space
limitation, penguin diagrams are not drawn in Fig. 3; they can be
obtained from Figs. 3(c)-3(g) by replacing the $b\to u$ tree
transition  by the $b\to s(d)$ penguin transition. Under the
factorization hypothesis, the relevant factorizable amplitudes are
proportional to $\la M|(\bar q_3 q_2)|0\ra\la \B_1\ov \B_2|(\bar
q_1b)|B\ra$ for Figs.~3(a) and 3(c), and to $\la \B_1\ov
\B_2|(\bar q_1 q_2)|0\ra \la M|(\bar q_3 b)|B\ra$ for Figs.~3(b)
and 3(d). For Figs. 3(b) and 3(d) the two-body matrix element $\la
\B_1\ov \B_2|(\bar q_1 q_2)|0\ra$ for octet baryons can be related
to the e.m. form factors of the nucleon. For Figs. 3(a) and 3(c)
we will consider the pole diagrams to evaluate 3-body matrix
elements. The 3-body matrix element $\la \B_1\ov \B_2|(\bar
q_1b)|B\ra$ receives contributions from point-like contact
interaction (i.e. direct weak transition) and pole diagrams
\cite{CKcharmless}.

\begin{figure}[t]
\includegraphics[width=7cm]{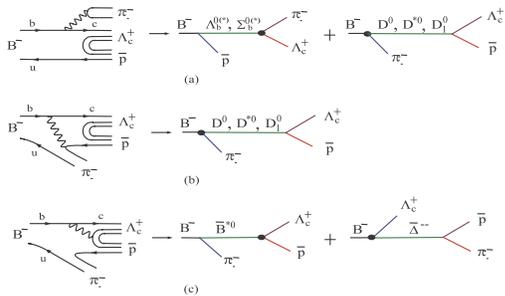}
\caption{Quark and pole diagrams for $B^-\to\Lambda_c^+\bar
    p\pi^-$, where the solid blob denotes the weak vertex. (a) and (b) correspond to
    factorizable external and internal $W$-emission
    contributions, respectively, while (c) to nonfactorizable internal $W$-emission
    diagrams.}
    \label{fig:Lamcppi}
\end{figure}

We consider the decay $B^-\to\Lambda_c^+\bar p\pi^-$ as an
illustration. It receives resonant and nonresonant contributions.
The factorizable nonresonant amplitude reads
 \be \label{factamp}
 A(B^-\to\Lambda_c^+\bar p\pi^-)_{\rm fact} &=&
 {G_F\over\sqrt{2}}V_{cb}V_{ud}^*\Big\{a_1\la \pi^-|(\bar
 du)|0\ra \non \\ && \times \la \Lambda_c^+\bar p|(\bar cb)|B^-\ra \non\\
 && +a_2\la\pi^-|(\bar db)|B^-\ra\la \Lambda_c^+\bar p|(\bar
 cu)|0\ra\Big\}\non \\ &\equiv&  A_1+A_2.
 \en
The factorizable amplitude $A_2$ can be directly calculated
\cite{CKcharm}. For the amplitude $A_1$ we evaluate the baryon and
meson pole diagrams in Fig. \ref{fig:Lamcppi}. Let's consider the
baryon pole due to $\Lambda_b$ in Fig. 4(a) whose propagator is
given by $1/( m_{\Lambda_b}^2-m_{\Lambda_c\pi}^2)$. In the heavy
quark limit, this propagator is not $1/m_b^2$ suppressed at the
region where the invariant mass of $\Lambda_c\pi$ pair is large,
for example, when the pion carries away much energies. For the
meson poles in Figs. 4(a) and 4(b), the $D$ meson propagator $ 1/(
m_D^2-m_{\Lambda_c \bar p}^2)$ is enhanced when the invariant mass
of $\Lambda_c\bar p$ is near the threshold. This is equivalent to
the $\Lambda_c\bar p$ form factor suppression at large momentum
transfer. Therefore, as far as the factorizable diagrams Figs.
4(a) and 4(b) are concerned, the most favorable configuration is
that the effective mass of $\Lambda_c\bar p$ is at its lowest
value. It should be stressed that this configuration is not
favored in Fig. 4(c). However, the latter contribution is
nonfactorizable and hence it is presumably suppressed. Note that
the measured spectrum can be used to constrain the momentum
dependence of baryon-pair form factors.

The pole diagram of Fig. 2(a) for two-body decays is always
$1/m_b^2$ suppressed and this explains why the three-body baryonic
$B$ decays can have rates larger than their two-body counterparts.

\begin{figure}[h]
\includegraphics[width=4cm]{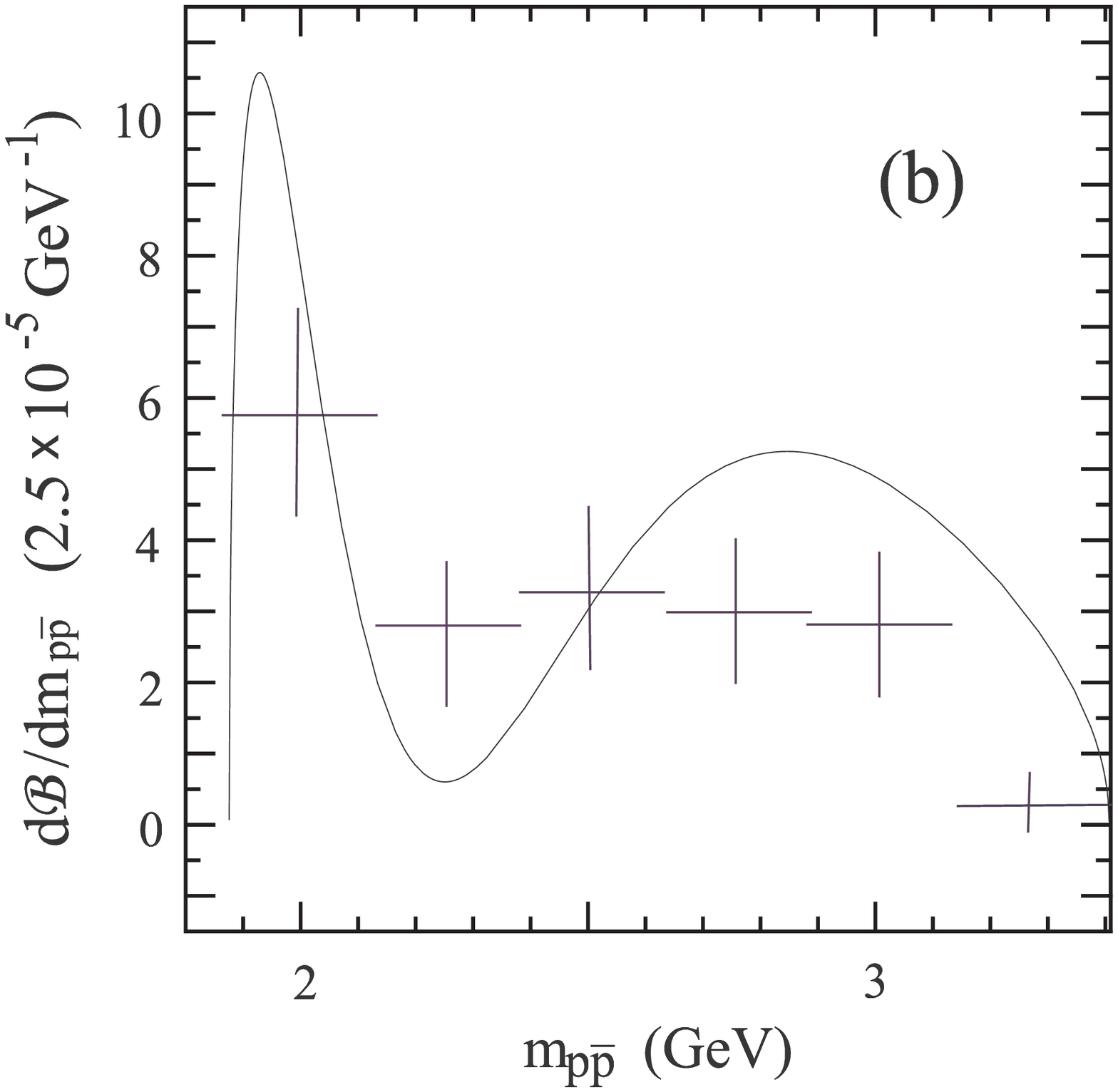}
\caption{The predicted $p\bar p$ invariant mass distribution of
$\ov B^0\to D^0p\bar p$ \cite{CKDmeson}. The experimental data are
taken from \cite{Belle:3charm}.} \label{fig:BDppinv}
\end{figure}

Many of 3-body baryonic $B$ decays have been studied in
\cite{CKcharm,CKcharmless,CKDmeson,CHT01,CHT02,CH02}. Because of
space limitation, we will focus on a few of the prominent features
of them:

\begin{itemize}
\item About 1/4 of the $B^-\to\Lambda_c^+\bar p\pi^-$ rate
 comes from resonant contributions \cite{CKcharm}.
\item Contrary to $\ov B^0\to D^{(*)+}n\bar p$ decays where the
$D^{*+}/D^+$ production ratio is anticipated to be of order 3, the
$D^{*0}/D^0$ production ratio in color suppressed $\ov B^0\to
D^{(*)0}p\bar p$ decays is consistent with unity experimentally
(see Table \ref{tab:3charm}). It is shown in \cite{CKDmeson} that
the similar rates for $D^0p\bar p$ and $D^{*0}p\bar p$ can be
understood within the framework of the pole model as the former is
dominated by the axial-vector meson states, whereas the other
modes proceed mainly through the vector meson poles.
 \item
The spectrum of $B\to D^0p\bar p$ is predicted to have a hump at
large $p\bar p$ invariant mass $m_{p\bar p}\sim 2.9$ GeV
\cite{CKDmeson} (see Fig. \ref{fig:BDppinv}), which needs to be
confirmed by forthcoming experiments.
 \item Charmless decays $B^-\to p\bar p
K^-(K^{*-})$ are penguin-dominated and have the branching ratios
 \be
 \B(B^-\to p\bar p K^-) &\approx& 4.0\times 10^{-6} ,  \non \\
 \B(B^-\to p\bar p K^{*-}) &\approx& 2.3\times 10^{-6},
 \en
predicted by the pole model \cite{CKcharmless}. It is naively
expected that $p\bar p K^{*-}<p\bar p K^-$ due to the absence of
$a_6$ and $a_8$ penguin terms to the former. The Belle observation
of a large rate for the $K^*$ production (see Table
\ref{tab:3charmless}) is thus unexpected. Also, it is non-trivial
to understand the observed sizable rate of $\ov B^0\to p\bar p \ov
K^0$ \cite{CKcharmless,CH02}.

 \item
The decay $B\to\Lambda\bar p\pi$ was previously argued to be small
($< 10^{-6}$) and suppressed relative to $\Sigma^0\bar p\pi^+$
\cite{CKcharmless}. The sizable branching ratio of the former
$\sim 4.0\times 10^{-6}$ observed recently by Belle
\cite{Belle:3charmless} can be understood now as a proper
treatment of the pseudoscalar form factor arising from penguin
matrix elements \cite{CH02}.

\end{itemize}

\section{Radiative baryonic $B$ decays}
Naively it appears that the bremsstrahlung process will lead to
$\Gamma(B\to\B_1\ov \B_2\gamma)\sim {\cal O}(\alpha_{\rm
em})\Gamma(B\to\B_1\ov \B_2)$ with $\alpha_{\rm em}$ being an
electromagnetic fine-structure constant and hence the radiative
baryonic $B$ decay is further suppressed than the two-body
counterpart, making its observation very difficult at the present
level of sensitivity for $B$ factories. However, there is an
important short-distance electromagnetic penguin transition $b\to
s \gamma$. Owing to the large top quark mass, the amplitude of
$b\to s\gamma$ is neither quark mixing nor loop suppressed.
Moreover, it is largely enhanced by QCD corrections. As a
consequence, the short-distance contribution due to the
electromagnetic penguin diagram dominates over the bremsstrahlung.
This phenomenon is quite unique to the bottom hadrons which
contain a heavy $b$ quark; such a magic short-distance enhancement
does not occur in the systems of charmed and strange hadrons.

Consider the $\Lambda_b$ pole diagram and apply heavy quark spin
symmetry and static $b$ quark limit to relate the tensor matrix
element appearing in
 \be
 && \la \Lambda(p_\Lambda)\gamma(\vp,k)|{\cal
 H}_W|\Lambda_b(p_{\Lambda_b})\ra = -i{G_F\over\sqrt{2}}\,{e\over 8\pi^2}
 V_{ts}^*V_{tb} \non \\  &&\qquad \times 2c_7^{\rm eff}
 m_b\vp^\mu k^\nu \la\Lambda|\bar
 s\sigma_{\mu\nu}(1+\gamma_5)b|\Lambda_b\ra
 \en
to the $\Lambda_b\to\Lambda$ form factors. The decay rate depends
on the strong coupling of $\Lambda_bB^+\bar p$ and
$\Lambda_b\to\Lambda$ transition form factors \cite{CKrad} and it
is found \footnote{The prediction of $B^-\to\Lambda\bar p\gamma$
in \cite{CKrad} is updated here using the strong coupling
constants constrained from the measured 3-body charmful baryonic
$B$ decays.}:
 \be
 \B(B^-\to\Lambda\bar p\gamma) &\approx&
 \B(B^-\to\Xi^0\bar\Sigma^-\gamma) \non \\
 &\sim&  1.2\times 10^{-6}.
 \en
Therefore, penguin-induced radiative baryonic $B$ decay modes
should be readily accessible by $B$ factories. Recently, CLEO
\cite{CLEOrad} has made the first attempt of measuring radiative
baryonic $B$ decays:
 \be
&& [\B(B^-\to\Lambda\bar p\gamma)+0.3\B(B^-\to\Sigma^0\bar
p\gamma)]_{E_\gamma>2.0\,{\rm GeV}} \non \\
 && \qquad <3.3\times 10^{-6}.
 \en
Theoretically, the production of $\Sigma^0\bar p\gamma$ is
suppressed relative to $\Lambda\bar p\gamma$ \cite{CKrad}.

\section{Conclusion}
Experimental and theoretical progresses in exclusive baryonic $B$
decays in the past few years are impressive. The threshold peaking
effect in baryon pair invariant mass is a key ingredient in
understanding three-body decays. The weak radiative baryonic
decays $B^-\to\Lambda\bar p\gamma$ and
$B^-\to\Xi^0\bar\Sigma^-\gamma$ mediated by the electromagnetic
penguin process $b\to s\gamma$ have branching ratios of order
$10^{-6}$ and should be readily accessible experimentally.

\begin{acknowledgments}
I wish to thank Kwei-Chou Yang for collaboration and Eung-Jin Chun
for organizing this wonderful conference.
\end{acknowledgments}


\end{document}